\documentclass [aps,amsmath,amssymb,superscriptaddress,preprint]{revtex4}
\usepackage{graphicx}
\usepackage{xcolor}

\def\gsim{\mathrel{\rlap{\lower4pt\hbox{\hskip1pt$\sim$}}
    \raise1pt\hbox{$>$}}}       

\def\be{\begin{equation}}
\def\bea{\begin{eqnarray}}
\def\ee{\end{equation}}
\def\eea{\end{eqnarray}}

\def\sq{{1\over \sqrt{2}}}

\def\z{\vert 0 \rangle}
\def\oo{\vert 1 \rangle}

\begin{document}
\title{Quantum Hair and Black Hole Information}
\author{Xavier Calmet} \email{x.calmet@sussex.ac.uk}
\affiliation{Department of Physics and Astronomy,\\
University of Sussex, Brighton, BN1 9QH, United Kingdom}

\author{Stephen~D.~H.~Hsu} \email{hsusteve@gmail.com}
\affiliation{Department of Physics and Astronomy\\ Michigan State University \\}

\begin{abstract}
It has been shown that the quantum state of the graviton field outside a black hole horizon carries information about the internal state of the hole. We explain how this allows unitary evaporation: the final radiation state is a complex superposition which depends linearly on the initial black hole state. Under time reversal, the radiation state evolves back to the original black hole quantum state. Formulations of the information paradox on a fixed semiclassical geometry describe only a small subset of the evaporation Hilbert space, and do not exclude overall unitarity.   
\end{abstract}

\maketitle
\section{Introduction}

In 1976 Stephen Hawking argued that black holes cause pure states to evolve into mixed states \cite{Hawking1976}. Put another way, quantum information that falls into a black hole does not escape in the form of radiation. Rather, it vanishes completely from our universe, thereby violating unitarity in quantum mechanics. 


Hawking’s arguments were based on the specific properties of black hole radiation. His calculations assumed a semiclassical spacetime background -- they did not treat spacetime itself in a quantum mechanical way, because this would require a theory of quantum gravity. The formulation of the information paradox has been refined over several decades, as briefly summarized below. See pedagogical reviews \cite{Mathur1,Mathur2}, and \cite{Raju:2020smc} for a recent and thorough discussion. 

\bigskip
\noindent
{\bf Hawking} (1976): Black hole radiation, calculated in a semiclassical spacetime background, is thermal and is in a mixed state. It therefore cannot encode the pure state quantum information behind the horizon. \cite{Hawking1976} 

\bigskip
\noindent
{\bf No Cloning} (circa 1990): There exist spacelike surfaces (``nice slices'')which intersect both the interior of the BH and the emitted Hawking radiation. The No Cloning theorem implies that the quantum state of the interior cannot be reproduced in the outgoing radiation. \cite{Preskill1992}

\bigskip
\noindent
{\bf Entanglement Monogamy} (circa 2010): Hawking modes are highly entangled with interior modes near the horizon, and therefore cannot purify the (late time) radiation state of an old black hole. \cite{Mathur1,Mathur2,firewall1,firewall2}

\bigskip
These formulations are limited by the assumption of a semiclassical spacetime background. Specifically, as we elaborate in what follows, they do not address the possibility of entanglement between different background geometries (gravitational states).

Recently it was shown \cite{QH} that the quantum state of the graviton field outside the horizon depends on the state of the interior. No-hair theorems in general relativity severely limit the information that can be encoded in the classical gravitational field of a black hole, but the situation is quite different at the quantum level. 

This result is directly connected to recent demonstrations \cite{R1,R2,R3} that the interior information is recoverable at the boundary: they originate, roughly speaking, from the Gauss Law constraint in quantization of gravity. It provides a mechanism (``quantum hair'') through which the information inside the hole is encoded in the quantum state of the exterior gravitational field.

\section{Quantum Hair and Radiation}

In \cite{QH}, analysis of the state of the graviton field produced by a compact matter source (e.g., a black hole) revealed the following. 

\bigskip

1. The asymptotic graviton state of an energy eigenstate source is determined at leading order by the energy eigenvalue. Insofar as there are no accidental energy degeneracies there is a one to one map between graviton states and matter source states. A semiclassical matter source produces an entangled graviton state.

2. Quantum gravitational fluctuations (i.e., graviton loops) produce corrections to the long range potential (e.g., $\sim r^{-5}$) whose coefficients depend on the internal state of the source. This provides an explicit example of how the graviton quantum state (corresponding to the semiclassical potential) encodes information about the internal state of a black hole.

\bigskip
Using property 1 above, we can write the quantum state of the exterior metric (equivalently, the quantum state of the exterior geometry) as 
\begin{equation}
 \Psi_i = \sum_n c_n \Psi_g (E_n) = \sum_n c_n \, \vert \, g(E_n) \, \rangle  ~.
\end{equation}
A semiclassical state has support concentrated in some range of energies $E$, where the magnitudes of $c_n$ are largest. For simplicity, when representing the exterior metric state $g(E)$ we only write the energy explicitly and suppress the other quantum numbers. In \cite{QH,Barnich,Mueck} the graviton state is given as a coherent state of Coulomb (non-propagating) graviton modes, which is a function of the energy eigenvalue of the matter source. However we do not require this explicit form to write an expression for the radiation state as a function of the black hole internal state.

Assume for convenience that the black hole emits one quantum at a time (e.g., at fixed intervals), culminating in a final state of $N$ radiation quanta:
$$ \vert ~ r_1 ~ r_2 ~ r_3 ~ \cdots ~ r_N ~ \rangle ~.$$
The quantum numbers of the $i$-th emitted radiation particle include the energy $\Delta_i$, momentum $p_i$, spin $s_i$, charge $q_i$, etc. The symbol $r_i$ is used to represent all of these values: $$r_i \sim \{ \Delta_i, p_i, s_i, q_i, \ldots \} ~.$$
A final radiation state is specified by the values of $\{r_1, r_2, \ldots , r_N \}$.

Let the amplitude for emission of quantum $r_i$ from exterior metric state $\Psi_g (E)$ be $\alpha (E,r_i)$. This amplitude must approximate the semiclassical Hawking amplitude for a black hole of mass $E$. In the leading approximation the amplitudes are those of thermal emission, but at subleading order (i.e., $\sim S^{-k}$ for perturbative corrections such as those calculated in \cite{QH}, or $\exp -S$ for nonperturbative effects, where $S$ is the black hole entropy) additional dependence on $(E, r_i)$ will emerge. The fact that these corrections can depend on the internal state of the hole is a consequence of quantum hair. It has been shown that even corrections as small as $\exp -S$ can purify a maximally mixed Hawking state (i.e., can perturb the radiation density matrix $\rho$ so that ${\rm tr} \, \rho^2 = 1$), because the dimensionality ($\sim \exp S$) of the Hilbert space is so large \cite{Raju:2020smc}.

\bigskip \noindent
When the black hole emits the first radiation quantum $r_1$ it evolves into the exterior state given on the right below:
\begin{equation}
    \Psi_i ~\rightarrow~ \sum_n \sum_{r_1} c_n ~ \alpha (E_n, r_1) ~ \vert \, g(E_n - \Delta_1), r_1 \rangle 
    \label{evolution1} ~~.
\end{equation}
In this notation $g$ refers to the exterior geometry and $r_1$ to the radiation. The next emission leads to
\begin{equation}
    \sum_n \sum_{r_1,r_2} c_n ~ \alpha (E_n, r_1) ~ \alpha (E_n - \Delta_1, r_2) ~
    \vert \, g(E_n - \Delta_1 - \Delta_2), r_1, r_2 \rangle 
    \label{evolution2} ~~,
\end{equation}
and the final radiation state is
\begin{equation}
\label{evolution3}
\sum_n \sum_{r_1,r_2,\ldots,r_N} c_n ~\alpha(E_n, r_1) ~\alpha(E_n - \Delta_1, r_2) ~ 
\alpha(E_n - \Delta_1 - \Delta_2, r_3 ) ~
\cdots ~ \vert \, r_1 ~ r_2 ~ \cdots ~ r_N \rangle ~. 
\end{equation}
In the final state we omit reference to the geometry $g$ as the black hole no longer exists: there is no horizon and the spacetime is approximately flat. 

As is the case with the decay of any macroscopic object into individual quanta, several remarks apply.
A semiclassical black hole state has support concentrated in a small range of energy eigenvalues $E_n$. Realistic radiation quanta, which are localized in space and time, are better represented as wave packet states than as plane wave states, and hence have a small spread as well in their energies $\Delta_i$. The energy conservation requirement: $E_n \approx \sum_i \, \Delta_i$ is not exact.

The final expression (\ref{evolution3}) is the same one we would obtain from a burning lump of coal if we interpret the amplitudes $\alpha (E, r)$ with $r$ (again) the state of the radiated particle and $E$ the state of the coal as it burns. 

Quantum hair allows the internal state of the black hole, reflected in the coefficients $c_n$, to affect the Hawking radiation. The result is manifestly unitary, and the final state in (\ref{evolution3}) is manifestly a pure state.

\bigskip \noindent

$\bullet$ For each distinct initial state given by the $\{ c_n \}$ there is a different final radiation state.

$\bullet$ The time-reversed evolution of a final radiation state results in a specific initial state.

\bigskip

Note the final radiation state in (\ref{evolution3}) is a macroscopic superposition state. Configurations described by $\{r_1, r_2, \ldots , r_N \}$ will exhibit different regions with higher or lower densities of energy, charge, etc. Time-reversed evolution will only produce the original semiclassical black hole state if these radiation states converge and interfere with the exact phase relations given in (\ref{evolution3}).

\section{Information Paradox Redux}

In this section we explain why the unitary evolution given in (\ref{evolution3}) is not excluded by previous arguments for an information paradox. We focus on an especially transparent formulation of the information paradox due to Mathur \cite{Mathur1,Mathur2}, later refined by others \cite{firewall1,firewall2}. This analysis tracks the entanglement entropy of Hawking radiation emitted on a nice slice.

The modes outside the horizon are labelled $b_1, b_2, ...$, and those inside the horizon are labelled $e_1, e_2, ...$ The initial slice in the nice slice foliation contains only the matter state $|\psi\rangle_M$ (i.e., the black hole), and none of the $e_i, b_i$. The first step of evolution stretches the spacelike slice, so that the particle modes $e_1, b_1$ are now present on the new slice. The state of these modes has the schematic form
\begin{equation}
\vert \Psi \rangle = \sq\Big ( \z_{e_1}\z_{b_1}+ \oo_{e_1}\oo_{b_1}\Big )~,
\label{three}
\end{equation}
where the numbers $0, 1$ give the occupation number of a particle mode. The entanglement of the state outside the horizon (given by the mode $b_1$) with the state inside  (given by the mode $e_1$) yields 
$S_{\rm entanglement}=\ln 2$.

In the next step of evolution the modes $b_1, e_1$  at the earlier step move apart, and in the region between them there appears another pair of modes $b_2, e_2$ in a state that has the same form as (\ref{three}). After $N$ steps we have the state
\begin{eqnarray}
\vert \Psi \rangle \approx ~ |\psi\rangle_M&\otimes&\Big( \sq \z_{e_1}\z_{b_1}+\sq\oo_{e_1}\oo_{b_1}\Big)\cr
&\otimes&\Big( \sq \z_{e_2} \z_{b_2} + \sq \oo_{e_2} \oo_{b_2} \Big)\cr
&\dots&\cr
&\otimes&\Big( \sq \z_{e_N} \z_{b_N} + \sq \oo_{e_N} \oo_{b_N} \Big)~.
\label{six}
\end{eqnarray}
The initial matter state appears in  a tensor product with all the other quanta, since the pair creation happens far away. There is no connection to the matter state in the leading order Hawking process, unlike in our (\ref{evolution3}) which is clearly non-factorizable.

The modes $\{ b_i\}$ are entangled with the  $\{M,  e_i\}$ with $S_{\rm entanglement}=N\ln 2$.
This entanglement grows by $\ln 2$ with each succeeding emission. If the hole evaporates away completely, the $b_i$ quanta outside will be in an entangled state, but there will be nothing that they are entangled {\it with}. The initial pure state $|\psi\rangle_M$ has evolved to a mixed state, described by a density matrix.

Mathur argues further that small corrections $\epsilon$ to Hawking evaporation cannot change this qualitative result: the entanglement entropy increases by $\approx \ln 2 - \epsilon$ with each emitted quantum.

However, the analysis is confined to a single semiclassical nice slice -- it reveals only a small portion of the evaporation process. Two radiation patterns $\{ b_1, b_2, ... , b_N \}$ and $\{ b'_1, b'_2, ... , b'_N \}$ which are sufficiently different correspond to distinct semiclassical geometries and hence distinct nice slices. This formulation of the paradox does not address the possibility of entanglement {\it between} different geometries. Furthermore, the nature of the infalling $e$ modes is different on each geometry as they are excitations with respect to a specific nice slice.

The $b$ modes are our $r$ modes. In our description the entanglement between $b$ and $e$ modes is reflected in the entanglement between $r$ modes and the exterior metric state $g(E)$, which is itself determined by the internal black hole state. A negative energy $e$ mode reduces the total energy of the internal state: $E \rightarrow E - \Delta$, and this is then reflected in the exterior metric state $g(E - \Delta)$. Other quantum numbers such as spin or charge which characterize the interior state are similarly modified by the infalling $e$ modes. 

The analog of the entangled state $\Big ( \z_{e_i}\z_{b_i}+ \oo_{e_i}\oo_{b_i}\Big )$ is simply
$$
\vert \, g(E - \Delta), r \, \rangle ~+~ \vert \, g(E - \Delta'), r' \, \rangle~~.
$$
Following Mathur, we have assumed that $r_i$ has only two possible states $r$ and $r'$ with energies $\Delta$ and $\Delta'$, respectively, and we have suppressed the other quantum numbers.
Thus we can see that the essential features of Mathur's nice slice analysis are contained in our expressions (\ref{evolution1}) -- (\ref{evolution3}). However there are two important differences: (A) In our treatment the $e$ modes modify the interior state, which then affects {\it subsequent} emissions via the exterior metric state. We do not simply trace over the $e$ modes as lost information: the entanglement between $b$ and $e$ re-emerges as entanglement between $b$ and subsequently emitted quanta. (B) Our expressions describe entanglement {\it across many} distinct geometries -- i.e., every geometry corresponding to each of the specific radiation states $\vert r_1 ~ r_2 ~ \cdots ~ r_N \rangle$. These additional correlations, which are not present in (\ref{six}), lead to a pure final radiation state (\ref{evolution3}). 

We note again that the $e$ modes themselves, and the argument that the Hawking radiation state has the form (\ref{three}) or (\ref{six}), are particular to a specific nice slice -- i.e., they originate from the vacuum state for that inertial frame. The entanglement entropy between $e$ and $b$ modes, which grows with each quantum emitted, is specific to the slice: it is {\it not} a global quantity. It cannot be, because the $e$ modes are not globally defined.

\bigskip
\bigskip

Prior formulations of the information paradox all have the same limitation, which we discuss below.
An argument for the information paradox must show that a black hole evaporates into a mixed final state, even if the initial state was pure. However, the Hilbert space of final states is extremely large, and analysis confined to a subset of the Hilbert space may indicate a mixed state because part of the (possibly pure) global state has been neglected -- equivalently, traced out.

Furthermore the final state is a superposition of many possible quantum spacetimes and corresponding radiation states: it is described by a wavefunction of the form  $\Psi [g, r]$  where $g$ describes the geometry and $r$ the radiation/matter fields. The entire black hole rest mass is eventually converted into radiation by the evaporation process, so fluctuations in the momenta of these radiation quanta can easily give the black hole a center of mass velocity which varies over the long evaporation time \cite{Hsu:2013cw,Hsu:2013fra}. The final spread in location of the black hole is of order the initial mass $M$ squared, so is much larger than its Schwarzschild radius. Each radiation pattern corresponds to a complex recoil trajectory of the black hole itself, and the resulting gravitational fields include exponentially many ($\sim \exp M^2$, modulo coarse graining) macroscopically distinct spacetimes \cite{Hsu:2013fra,Bao:2017who}.

Therefore, restriction to a specific semiclassical background metric (i.e., nice slice) is a restriction to a very small subset $X$ of the final state Hilbert space $Y$. Concentration of measure (Levy's lemma) \cite{Concentration,Buniy:2020dux} implies that for almost all pure states $\Psi$ in a large Hilbert space $Y$, the density matrix 
$$
 \rho (X) =  {\rm tr} ~\Psi^\dagger \, \Psi 
$$
describing a small subset $X$ (tracing over the complement of $X$ in $Y$) is nearly maximally mixed -- i.e., like the radiation found in Hawking's original calculation \cite{Hsu:2009ve,Hsu:2010sb}.

Analysis restricted to a specific spacetime background is only sensitive to the subset $X$ of Hilbert space consistent with that semiclassical description. The analysis only probes the mixed state $\rho (X)$ and not the entire state $\Psi$ which exists in the large Hilbert space $Y$. Thus even if the evaporation is entirely unitary, resulting in a pure final state $\Psi$ in $Y$, it might {\it appear} to violate unitarity if the analysis is restricted to $X$ and hence only investigates the mixed state $\rho (X)$. 

In other words, a calculation performed on a specific nice slice might indicate that pure black hole states {\it seem} to evolve to mixed states. But the result does not address global unitarity. Entanglement between different $X$ and $X'$ -- equivalently, between different branches of the wavefunction $\Psi$ -- has been neglected, although even exponentially small entanglement between these branches may be sufficient to unitarize the result \cite{Hsu:2013fra,Raju:2020smc}. The final state (\ref{evolution3}) exhibits such $X$-$X'$ entanglements. See our earlier discussion of the size of deviations of amplitudes $\alpha (E, r)$ from thermal Hawking radiation.

\section{Conclusion}

Hawking's information paradox has been the focus of intense interest for almost 50 years. In his 1992 lecture on the subject, John Preskill wrote \cite{Preskill1992}

\bigskip \noindent
{\it  I conclude that the information loss paradox may well presage a revolution in fundamental physics.}
\bigskip

The resolution described here is conservative: the quantum state of the exterior gravity field is determined by the interior black hole state, allowing the latter to influence Hawking radiation produced at the horizon. Two distinct quantum states of the black hole may produce the same semiclassical external geometry, but the graviton states differ at the quantum level. The relationship between interior and exterior quantum states is not governed by classical no-hair theorems. Indeed, it has gradually been appreciated that gravity itself prevents the localization of quantum information \cite{Marolf:2008mf,R1,R2,R3,Raju:2020smc,Jacobson:2012ubm,Jacobson:2019gnm}, even behind a horizon. We stress that all formulations of the paradox require a degree of factorization between the black hole internal state and the radiation (see, e.g., (\ref{six})), which is clearly not true of our equation (\ref{evolution3}).

Certain aspects of our expressions (\ref{evolution1})-(\ref{evolution3}) are very clear: the black hole information is spread over many branches of the final radiation state, and macroscopic superpositions of different spacetime geometries play a role in the evaporation. Some of the difficulty in resolving the paradox may originate from a reluctance to accept these aspects of quantum dynamics.

\bigskip

{\it Acknowledgments:}
The authors thank R. Casadio, F. Kuipers, S. Raju, and S. Mathur for discussions. The work of X.C. is supported in part  by the Science and Technology Facilities Council (grants numbers ST/T00102X/1 and ST/T006048/1).

\end{document}